\documentclass[prl,preprint,12pt,onecolumn]{revtex4}
\usepackage{amsmath}

\input{tcilatex}

\begin{document}

\title{Practical creation and detection of polarization Bell states using
parametric down-conversion}
\author{K. J. Resch, J. S. Lundeen, and A. M. Steinberg}
\address{Department of Physics, University of Toronto\\
60 St. George Street, Toronto ON M5S 1A7\\
CANADA}

\begin{abstract}
The generation and detection of maximally-entangled two-particle states,
`Bell states,' are crucial tasks in many quantum information protocols such
as cryptography, teleportation, and dense coding. \ Unfortunately, they
require strong inter-particle interactions lacking in optics. \ For this
reason, it has not previously been possible to perform complete Bell state
determination in optical systems. \ In this work, we show how a recently
developed quantum interference technique for enhancing optical
nonlinearities can make efficient Bell state measurement possible. \ We also
discuss weaknesses of the scheme including why it cannot be used for
unconditional quantum teleportation.
\end{abstract}

\maketitle

\section{Introduction\qquad}

The new science of quantum information builds on the recognition that
entanglement, an essential but long underemphasized feature of quantum
mechanics, can be a valuable resource. \ Many of the headline-grabbing
quantum communication schemes (including quantum teleportation \cite%
{bennbrass,zeil,demartini}, dense coding \cite{densetheory,denseexp}, and
quantum cryptography \cite{cryptbennbrass,cryptekert}) are based on the
maximally-entangled two-particle quantum states called Bell states. \ Using
the polarization states of a pair of photons in different spatial modes, the
four Bell states are written as:%
\begin{eqnarray}
\left| \psi ^{\pm }\right\rangle &=&\frac{1}{\sqrt{2}}\left( \left|
V\right\rangle _{1}\left| H\right\rangle _{2}\pm \left| H\right\rangle
_{1}\left| V\right\rangle _{2}\right)  \notag \\
\left| \phi ^{\pm }\right\rangle &=&\frac{1}{\sqrt{2}}\left( \left|
H\right\rangle _{1}\left| H\right\rangle _{2}\pm \left| V\right\rangle
_{1}\left| V\right\rangle _{2}\right) ,
\end{eqnarray}%
where $\left| H\right\rangle $ and $\left| V\right\rangle $ describe
horizontal- and vertical-polarization states, and the subscripts 1 and 2 are
spatial mode labels. \ These four states form a complete, orthonormal basis
for the polarization states of a pair of photons. \ In each Bell state, a
given photon is completely unpolarized but perfectly correlated with the
polarization of the other photon. \ Photon Bell states were produced in
atomic cascades for the first tests of the nonlocal predictions of quantum
mechanics \cite{clauserbell}. \ Since that time, parametric down-conversion
sources \cite{othersources,steinberg,kwiat1,kwiat2,shapiro} have replaced
cascade souces due to their ease of use, high brightness, and the
high-purity states they produce. \ However, down-conversion sources do not
deterministically prepare photon Bell states, but rather states in which the
Bell state component is in a coherent superposition with a dominant vacuum
term; coincidence detection of photon pairs projects out only the two-photon
component of the state.

While optical Bell state source technology has shown marked improvement,
methods of distinguishing these states has proven a difficult challenge. \
Perhaps the most well-known example of why distinguishing Bell states is
important comes from quantum teleportation. \ A general projective
measurement is required for unconditional teleportation; experimental
teleportation was originally limited to a maximum efficiency of 25\% since
only the singlet state, $\left| \psi ^{-}\right\rangle ,$ could be
distinguished from the triplet states \cite{zeil}. \ The challenge for
measuring Bell states stems from the requirement for a strong inter-particle
interaction, which is usually nonexistent for photons. \ Without such a
nonlinearity, only two of the four states can be distinguished\cite%
{circuits2}. \ It was realized that a strong enough optical nonlinearity,
typically $\chi ^{(3)}$, could be used to mediate a photon-photon
interaction. \ Unfortunately, even the nonlinearities of our best materials
are far too weak. \ An experiment using standard nonlinear materials to
demonstrate a scheme for unconditional teleportation was limited to
extremely low efficiencies (on the order of 10$^{-10}$) by the tiny
nonlinearities involved \cite{shihtelep}. \ Proposals for extending optical
nonlinearities to the quantum level include schemes based on cavity QED \cite%
{qed}, electromagnetically-induced transparency \cite{eit}, photon-exchange
interactions \cite{pei}, and quantum interference techniques\ \cite%
{switch,phase}. \ Using the latter, we have recently demonstrated a
conditional-phase switch \cite{phase} which is similar to the
controlled-phase gate in quantum computation. \ In this work, we show how to
apply the conditional-phase switch to the problem of Bell state detection. \
It should be noted that if recently published schemes for performing quantum
computing with linear optics \cite{KLM,otherlin} could be experimentally
realized, then the problem of distinguishing all four Bell states could be
performed without the need for strong optical nonlinearities. \ Theoretical
work has also shown that if the Bell state is embedded appropriately in a
higher-dimensional Hilbert space, all of the Bell states can be
distinguished \cite{embedded}.

Strong optical nonlinearities are desired so that one can construct a
controlled-$\pi $, a specific case of the controlled-phase gate for photons.
\ Such a gate and all one-qubit rotations form a universal set of gates for
the more general problem of quantum computation -- just as the NAND gate is
universal for classical computation. \ The controlled-$\pi $ transformation %
\cite{cphiref} is described by:%
\begin{eqnarray}
\left| 0\right\rangle _{1}\left| 0\right\rangle _{2} &\longrightarrow
&\left| 0\right\rangle _{1}\left| 0\right\rangle _{2}  \notag \\
\left| 0\right\rangle _{1}\left| 1\right\rangle _{2} &\longrightarrow
&\left| 0\right\rangle _{1}\left| 1\right\rangle _{2}  \notag \\
\left| 1\right\rangle _{1}\left| 0\right\rangle _{2} &\longrightarrow
&\left| 1\right\rangle _{1}\left| 0\right\rangle _{2}  \notag \\
\left| 1\right\rangle _{1}\left| 1\right\rangle _{2} &\longrightarrow
&-\left| 1\right\rangle _{1}\left| 1\right\rangle _{2},  \label{cpi}
\end{eqnarray}%
in which the two qubit states are $\left| 0\right\rangle $ and $\left|
1\right\rangle $ and the subscript is the qubit label. \ This transformation
does nothing to the input state unless both qubits have a value of $\left|
1\right\rangle ,$ in which case it applies a phase-shift of $\pi .$ \ On the
surface this transformation appears to do nothing since an overall phase in
quantum mechanics is meaningless. \ However, it is clearly nontrivial when
applied to superpositions of states. \ 

The polarization of the photon makes an ideal two-level system for encoding
a qubit largely due to its relative immunity to environmental decoherence. \
A\ large enough $\chi ^{(3)}$ nonlinearity could be used to effect the c-$%
\pi $ transformation on a pair of photons. \ Given a polarization-dependent $%
\chi ^{(3)}$, or through the use of polarizing beam-splitters, only photon
pairs with, say, horizontal polarization would experience the nonlinear
interaction and pick up the additional phase shift. \ Such a gate could then
be incorporated into the optical implementation of the quantum circuits
shown in Fig. 1a. and 2a. (similar circuits are discussed in \cite%
{circuits2,circuits}). \ The circuit in Fig. 1a. converts, through unitary
transformation, a state in the rectilinear product state basis (i.e. $\left|
0\right\rangle _{1}\left| 0\right\rangle _{2},\left| 0\right\rangle
_{1}\left| 1\right\rangle _{2},\left| 1\right\rangle _{1}\left|
0\right\rangle _{2},$ and $\left| 1\right\rangle _{1}\left| 1\right\rangle
_{2}$) to the Bell basis. \ The circuit in Fig. 2a. performs the opposite
function converting a Bell state via unitary transformation to the
rectilinear basis. \ In essence, these circuits allow for the creation and
removal of entanglement between pairs of qubits. \ If the qubit states $%
\left| 0\right\rangle $ and $\left| 1\right\rangle $ are encoded into the
polarization states $\left| H\right\rangle $ and $\left| V\right\rangle $ in
two different spatial modes 1 and 2, then an optical realization of the
circuit in Fig. 2a. allows for the conversion of a photon pair in a Bell
state to a rectilinear basis state. \ These four rectilinear basis states
are easily distinguishable using the simple optical setup shown in Fig. 3. \
Thus, after passing the photon pair in a Bell state through the optical
realization of the circuit in Fig. 2a., the subsequent detection of the
rectilinear state is equivalent to determination of the Bell state.

The conditional-phase switch we propose is related to the controlled-phase
gate of quantum computation and is described in the theory section of this
work. \ The switching effect occurs in a $\chi ^{(2)}$ nonlinear material
that is pumped by a strong, classical beam. \ This pump beam is capable of
creating pairs of down-converted photon pairs into a pair of output modes. \
Pairs of photons, in a coherent superposition with the vacuum, pass through
the crystal into those same output modes. \ It is the interference between
the amplitudes for multiple paths leading to a photon pair that greatly
enhances the effective nonlinearity; since the down-converted light is only
created in pairs, the interference only affects the amplitude for photon
pairs. \ However, since the switching effect is based on an interference
effect, it is intrinsically dependent on the phase and amplitude of the
incoming beams. \ This has two consequences. \ First, the switch requires an
input which is in a coherent superposition with the vacuum. \ In this way,
the input has the required \emph{uncertain} number of photons, since photon
number and phase are conjugate quantities. \ And second, the switch works as
described only for states in the correct superposition with the vacuum, not
a general input state. \ As we will show, these conditions do allow for one
to distinguish between the four Bell states provided they are in the correct
superposition with the vacuum. \ Nonetheless, the conditions are too
stringent to allow for unconditional teleportation using this method.

First, we describe the effective nonlinearity. \ Then we show how the
nonlinearity can be used to construct optical devices analogous to the
quantum computation circuits shown in Fig. 1a. and Fig. 2a. \ 

\section{Theory}

\subsection{Effective Nonlinearity}

The general down-conversion state can be written as%
\begin{equation}
\left| \psi \right\rangle =\left| 0\right\rangle +\varepsilon \left( 
\begin{array}{cccc}
\left| H\right\rangle _{1}\left| H\right\rangle _{2}\text{ \ \ } & \left|
H\right\rangle _{1}\left| V\right\rangle _{2}\text{ \ \ } & \left|
V\right\rangle _{1}\left| H\right\rangle _{2}\text{ \ \ } & \left|
V\right\rangle _{1}\left| V\right\rangle _{2}%
\end{array}%
\right) \left( 
\begin{array}{c}
\alpha \\ 
\beta \\ 
\gamma \\ 
\delta%
\end{array}%
\right) ,  \label{dcstate}
\end{equation}%
where the part of the state describing photon pairs has been written as an
inner product. \ The amplitudes for the polarization states $\left|
H\right\rangle _{1}\left| H\right\rangle _{2}$, $\left| H\right\rangle
_{1}\left| V\right\rangle _{2}$, $\left| V\right\rangle _{1}\left|
H\right\rangle _{2}$, and $\left| V\right\rangle _{1}\left| V\right\rangle
_{2}$ are $\varepsilon \alpha $, $\varepsilon \beta ,$ $\varepsilon \gamma $%
, and $\varepsilon \delta $, respectively. \ Again, the subscripts 1 and 2
describe two different spatial modes. \ Throughout this theory section, we
adopt a 4-dimensional vector representation to describe the polarization
state of the photon pairs. \ In this more compact notation, the general
state is written 
\begin{equation}
\left| \psi \right\rangle =\left| 0\right\rangle +\varepsilon \left( 
\begin{array}{c}
\alpha \\ 
\beta \\ 
\gamma \\ 
\delta%
\end{array}%
\right) ,
\end{equation}%
In both cases, we have suppressed the normalization factor for clarity, and
for the discussion here we will restrict ourselves to the case where the
probability of having a photon pair at any given time is small, i.e. $\left|
\varepsilon \right| ^{2}\ll 1$ (as is always the case in real
down-conversion experiments).

The effective nonlinearity \cite{phase} can be described as follows. \ Modes
1 and 2 are of frequency $\omega $ and pass through a $\chi ^{(2)}$
nonlinear crystal that is simultaneously pumped by a strong classical laser
beam of frequency $2\omega $ in mode p. \ The modes are so chosen such that
the nonlinear crystal can create degenerate horizontally-polarized photon
pairs in spatial modes 1 and 2 via spontaneous parametric down-conversion,
as shown in Fig. 4. \ The nonlinear process is mediated by the interaction
Hamiltonian,

\begin{equation}
\mathcal{H}=ga_{1,H}^{\dagger }a_{2,H}^{\dagger }a_{p,V}+g^{\ast
}a_{1,H}a_{2,H}a_{p,V}^{\dagger },
\end{equation}%
where g is the coupling constant and $a_{i}^{(\dagger )}$ is the field
annihilation (creation) operator for the $i^{th}$ mode, and the subscripts $%
H $ and $V$ are the polarizations of the relevant modes for the type-I\
phase-matching. \ The pump laser is intense enough that we treat it
classically by replacing its field operators with c-number amplitudes, $%
\zeta $ and $\zeta ^{\ast }$:%
\begin{equation}
\mathcal{H}=g\zeta a_{1,H}^{\dagger }a_{2,H}^{\dagger }+g^{\ast }\zeta
^{\ast }a_{1,H}a_{2,H}.
\end{equation}%
Due to phase-matching constraints, the nonlinear crystal can only produce
horizontally-polarized photon pairs. \ In the weak coupling regime, we can
use first-order perturbation theory to propagate our state under the
interaction to,%
\begin{eqnarray}
\left| \psi (t)\right\rangle &=&\left( 1-\frac{it}{\hbar }\mathcal{H}\right)
\left| \psi \right\rangle \\
&=&\left| 0\right\rangle +\varepsilon \left( 
\begin{array}{c}
\alpha \\ 
\beta \\ 
\gamma \\ 
\delta%
\end{array}%
\right) -\frac{it}{\hbar }g\zeta \left( 
\begin{array}{c}
1 \\ 
0 \\ 
0 \\ 
0%
\end{array}%
\right) \\
&=&\left| 0\right\rangle +\varepsilon \left( 
\begin{array}{c}
\alpha -\frac{it}{\hbar }\frac{g\zeta }{\varepsilon } \\ 
\beta \\ 
\gamma \\ 
\delta%
\end{array}%
\right) .
\end{eqnarray}%
To first order, this Hamiltonian simply creates an amplitude for a
horizontally-polarized pair of photons. \ This new down-conversion amplitude
interferes with the preexisting amplitude for the $HH$ term. \ 

The transformation, as described here, does not appear unitary. \ This is
due to a few approximations. \ We assume that the vacuum term in our state
is unchanged, and neglect terms describing more that one pair of photons. \
These approximations are only valid in the relevant limit where $\left|
\varepsilon \right| $ $\ll 1$, where we can also suppress the normalization
term for clarity. \ However, the exact propagator follows from a hermitian
Hamiltonian and is of course unitary.

As was shown in the ``railcross experiment'' \cite{railcross} and in our
subsequent work with photon pairs from coherent state inputs \cite{switch},
interference between the amplitudes for existing pairs and for
down-conversion can modulate the rate of pair production. \ Given the
phase-matching scheme presented here, only the amplitude for $HH$ pairs is
affected. \ Accompanying this modulation of the photon pair production rate
is a shift in the phase of the horizontally-polarized photon pair term. \
The down-conversion crystal impresses a $\pi $ phase-shift on the $HH$ term
if the down-conversion amplitude, $-itg\zeta /\hbar $ to be $-2\varepsilon
\alpha $. \ To implement a tranformation analogous to the c-$\pi $ (Eq. \ref%
{cpi}) in the coincidence basis, this is the only condition that must be
enforced; the values for the coefficients $\alpha $, $\beta $, and $\gamma $
are free. \ This condition takes the place of the more usual normalization
condition on $\alpha $, $\beta $, $\gamma ,$ and $\delta $ to describe our
state space. \ It can be enforced experimentally by controlling the
amplitude and phase of the pump laser and/or the overall pair amplitude $%
\varepsilon $. \ Unfortunately, this means that the gate cannot be utilized
on arbitrary inputs without some prior information. \ Under these
conditions, the crystal implements 
\begin{equation}
\left| 0\right\rangle +\varepsilon \left( 
\begin{array}{c}
\alpha \\ 
\beta \\ 
\gamma \\ 
\delta%
\end{array}%
\right) \longrightarrow \left| 0\right\rangle +\varepsilon \left( 
\begin{array}{c}
-\alpha \\ 
\beta \\ 
\gamma \\ 
\delta%
\end{array}%
\right) .
\end{equation}%
If horizontal polarization is used to represent a logical `0', this performs
a transformation analogous to a c-$\pi $ within the state space defined by
our constraint on $\alpha .$ \ We do not use the conventional c-$\pi $ so
that we can use the common convention for the Hadamard gate later on without
the need for additional quantum gates. \ We will now describe how this
operation can be used to perform Bell state creation under certain
conditions. \ 

\subsection{Bell state creation}

The circuit in Fig. 1a. is capable of converting each rectilinear basis
state to a different Bell state. \ To give a concrete example, we begin with
the qubit pair in the state $\left| 0\right\rangle _{1}\left| 0\right\rangle
_{2}\ $represented as the 4-vector%
\begin{equation}
\left| \psi \right\rangle =\left( 
\begin{array}{c}
1 \\ 
0 \\ 
0 \\ 
0%
\end{array}%
\right) ,
\end{equation}%
where the rows now contain the amplitudes for the states $\left|
0\right\rangle _{1}\left| 0\right\rangle _{2}$, $\left| 0\right\rangle
_{1}\left| 1\right\rangle _{2}$, $\left| 1\right\rangle _{1}\left|
0\right\rangle _{2}$, and $\left| 1\right\rangle _{1}\left| 1\right\rangle
_{2}.$ \ The circuit contains one-qubit Hadamard transformations which are
defined by the 2$\times $2 matrix,

\begin{equation}
H=\frac{1}{\sqrt{2}}\left[ 
\begin{array}{cc}
1 & 1 \\ 
1 & -1%
\end{array}%
\right]
\end{equation}%
and the two-qubit c-$\pi $ gate whose operation has already been discussed.
\ The circuit then takes the input state, $\left| \psi \right\rangle $, to
the output state $\left| \psi ^{\prime }\right\rangle $ given by%
\begin{eqnarray}
\left| \psi ^{\prime }\right\rangle &=&\left( H_{1}\otimes I_{2}\right)
\left( c\text{-}\pi \right) \left( H_{1}\otimes H_{2}\right) \left| \psi
\right\rangle \\
&=&\frac{1}{2\sqrt{2}}\left[ 
\begin{array}{cccc}
1 & 0 & 1 & 0 \\ 
0 & 1 & 0 & 1 \\ 
1 & 0 & -1 & 0 \\ 
0 & 1 & 0 & -1%
\end{array}%
\right] \left[ 
\begin{array}{cccc}
1 & 0 & 0 & 0 \\ 
0 & 1 & 0 & 0 \\ 
0 & 0 & 1 & 0 \\ 
0 & 0 & 0 & -1%
\end{array}%
\right] \left[ 
\begin{array}{cccc}
1 & 1 & 1 & 1 \\ 
1 & -1 & 1 & -1 \\ 
1 & 1 & -1 & -1 \\ 
1 & -1 & -1 & 1%
\end{array}%
\right] \left( 
\begin{array}{c}
1 \\ 
0 \\ 
0 \\ 
0%
\end{array}%
\right) \\
&=&\frac{1}{\sqrt{2}}\left( 
\begin{array}{c}
1 \\ 
0 \\ 
0 \\ 
1%
\end{array}%
\right) .
\end{eqnarray}%
This final state is the Bell state $\left| \phi ^{+}\right\rangle $. \ Each
different rectilinear state input will produce a different Bell state output
through this circuit.

The conditional-phase operation can be incorporated into the optical device
schematically represented in Fig. 1b that can perform a very similar
transformation. \ Instead of using a state describing a pure photon pair as
input, this device requires the input pair to be in a coherent superposition
with the vacuum. \ As discussed previously, this is merely the output from a
parametric down-conversion source (Eq. \ref{dcstate}). \ Here we assume the
coefficients are normalized according to $\left| \alpha \right| ^{2}+\left|
\beta \right| ^{2}+\left| \gamma \right| ^{2}+\left| \delta \right| ^{2}=1,$
such that $\left| \varepsilon \right| ^{2}$ is the probability of a photon
pair of any polarization being present. \ The photons have been created into
spatial modes 1 and 2 by an initial down-conversion crystal (not shown) to
serve as input to the optical device in Fig. 1b. \ Hadamard operations are
accomplished via half-wave plates at 22.5 degrees, and the c-$\pi $ has been
replaced by the conditional-phase switch. \ The intial state will evolve as
follows through the device. \ The pair of Hadamard gates changes the general
state, $\left| \psi _{1}\right\rangle ,$ to $\left| \psi _{2}\right\rangle $%
, 
\begin{eqnarray}
\left| \psi _{2}\right\rangle  &=&\left( H_{1}\otimes H_{2}\right) \left|
\psi _{1}\right\rangle  \\
&=&\left| 0\right\rangle +\frac{\varepsilon }{2}\left[ 
\begin{array}{cccc}
1 & 1 & 1 & 1 \\ 
1 & -1 & 1 & -1 \\ 
1 & 1 & -1 & -1 \\ 
1 & -1 & -1 & 1%
\end{array}%
\right] \left( 
\begin{array}{c}
\alpha  \\ 
\beta  \\ 
\gamma  \\ 
\delta 
\end{array}%
\right)  \\
&=&\left| 0\right\rangle +\frac{\varepsilon }{2}\left( 
\begin{array}{c}
\alpha +\beta +\gamma +\delta  \\ 
\alpha -\beta +\gamma -\delta  \\ 
\alpha +\beta -\gamma -\delta  \\ 
\alpha -\beta -\gamma +\delta 
\end{array}%
\right) .
\end{eqnarray}%
This state passes through the conditional-phase shift, which is
phase-matched to contribute an amplitude of $-\varepsilon $ for
horizontally-polarized photon pairs. \ It will evolve to $\left| \psi
_{3}\right\rangle ,$%
\begin{eqnarray}
\left| \psi _{3}\right\rangle  &=&\left| 0\right\rangle +\frac{\varepsilon }{%
2}\left( 
\begin{array}{c}
\alpha +\beta +\gamma +\delta  \\ 
\alpha -\beta +\gamma -\delta  \\ 
\alpha +\beta -\gamma -\delta  \\ 
\alpha -\beta -\gamma +\delta 
\end{array}%
\right) -\varepsilon \left( 
\begin{array}{c}
1 \\ 
0 \\ 
0 \\ 
0%
\end{array}%
\right)  \\
&=&\left| 0\right\rangle +\frac{\varepsilon }{2}\left( 
\begin{array}{c}
\alpha +\beta +\gamma +\delta -2 \\ 
\alpha -\beta +\gamma -\delta  \\ 
\alpha +\beta -\gamma -\delta  \\ 
\alpha -\beta -\gamma +\delta 
\end{array}%
\right) .
\end{eqnarray}%
The final Hadamard gate acts only on mode 1, and converts $\left| \psi
_{3}\right\rangle $ to the output state $\left| \psi ^{\prime }\right\rangle 
$,

\begin{eqnarray}
\left| \psi ^{\prime }\right\rangle &=&\left( H_{1}\otimes I_{2}\right)
\left| \psi _{3}\right\rangle \\
&=&\left| 0\right\rangle +\frac{\varepsilon }{2\sqrt{2}}\left[ 
\begin{array}{cccc}
1 & 0 & 1 & 0 \\ 
0 & 1 & 0 & 1 \\ 
1 & 0 & -1 & 0 \\ 
0 & 1 & 0 & -1%
\end{array}%
\right] \left( 
\begin{array}{c}
\alpha +\beta +\gamma +\delta -2 \\ 
\alpha -\beta +\gamma -\delta \\ 
\alpha +\beta -\gamma -\delta \\ 
\alpha -\beta -\gamma +\delta%
\end{array}%
\right) \\
&=&\left| 0\right\rangle +\frac{\varepsilon }{\sqrt{2}}\left( 
\begin{array}{c}
\alpha +\beta -1 \\ 
\alpha -\beta \\ 
\gamma +\delta -1 \\ 
\gamma -\delta%
\end{array}%
\right) .
\end{eqnarray}

If, for example, the input state to this device had only an amplitude for a
horizontally-polarized photon pair (i.e. $\alpha =1$ and $\beta ,\gamma
,\delta =0$), then the output state would be,%
\begin{eqnarray}
\left| \psi ^{\prime }\right\rangle &=&\left| 0\right\rangle +\frac{%
\varepsilon }{\sqrt{2}}\left( 
\begin{array}{c}
0 \\ 
1 \\ 
-1 \\ 
0%
\end{array}%
\right) \\
&=&\left| 0\right\rangle -\varepsilon \left| \psi ^{-}\right\rangle .
\end{eqnarray}%
The other 3 possible rectilinear basis inputs would each evolve to a
different Bell state in a coherent superposition with the vacuum state. \
The resulting transformations on four possible rectilinear input states are
\ 
\begin{eqnarray}
\left| 0\right\rangle +\varepsilon \left| H\right\rangle _{1}\left|
H\right\rangle _{2} &\longrightarrow &\left| 0\right\rangle -\varepsilon
\left| \psi ^{-}\right\rangle  \notag \\
\left| 0\right\rangle +\varepsilon \left| H\right\rangle _{1}\left|
V\right\rangle _{2} &\longrightarrow &\left| 0\right\rangle -\varepsilon
\left| \psi ^{+}\right\rangle  \notag \\
\left| 0\right\rangle +\varepsilon \left| V\right\rangle _{1}\left|
H\right\rangle _{2} &\longrightarrow &\left| 0\right\rangle -\varepsilon
\left| \phi ^{-}\right\rangle  \notag \\
\left| 0\right\rangle +\varepsilon \left| V\right\rangle _{1}\left|
V\right\rangle _{2} &\longrightarrow &\left| 0\right\rangle -\varepsilon
\left| \phi ^{+}\right\rangle .
\end{eqnarray}%
\ 

\subsection{Bell state detection}

The method just described for creating polarization Bell states is much more
experimentally difficult than the elegant methods of doing so in a
cleverly-oriented crystal or crystal pair \cite{kwiat1,kwiat2}. \ What is
unique about this method is that this device performs a one-to-one
transformation between rectilinear basis states and Bell basis states. \
This device for creating the Bell states can, in fact, be run in reverse to
distinguish between the four Bell states provided, again, that they are in a
superposition with vacuum. \ Fig. 2a. shows a quantum circuit for
transforming Bell states to the rectilinear basis, that is very similar in
structure to the circuit shown in Fig. 1a. \ To give a concrete example, we
can trace the evolution of the singlet state, $\left| \phi ^{-}\right\rangle 
$, through the device. \ The singlet state can be written in 4-vector
notation as,%
\begin{equation}
\left| \psi ^{-}\right\rangle =\frac{1}{\sqrt{2}}\left( 
\begin{array}{c}
0 \\ 
-1 \\ 
1 \\ 
0%
\end{array}%
\right) .
\end{equation}%
The circuit transforms the input state to the output $\left| \psi ^{\prime
}\right\rangle $ in the following way,%
\begin{eqnarray}
\left| \psi ^{\prime }\right\rangle  &=&\left( H_{1}\otimes H_{2}\right)
\left( c\text{-}\pi \right) \left( H_{1}\otimes I_{2}\right) \left| \phi
^{-}\right\rangle  \\
&=&\frac{1}{2}\left[ 
\begin{array}{cccc}
1 & 1 & 1 & 1 \\ 
1 & -1 & 1 & -1 \\ 
1 & 1 & -1 & -1 \\ 
1 & -1 & -1 & 1%
\end{array}%
\right] \left[ 
\begin{array}{cccc}
1 & 0 & 0 & 0 \\ 
0 & 1 & 0 & 0 \\ 
0 & 0 & 1 & 0 \\ 
0 & 0 & 0 & -1%
\end{array}%
\right] \frac{1}{\sqrt{2}}\left[ 
\begin{array}{cccc}
1 & 0 & 1 & 0 \\ 
0 & 1 & 0 & 1 \\ 
1 & 0 & -1 & 0 \\ 
0 & 1 & 0 & -1%
\end{array}%
\right] \frac{1}{\sqrt{2}}\left( 
\begin{array}{c}
0 \\ 
-1 \\ 
1 \\ 
0%
\end{array}%
\right)  \\
&=&\left( 
\begin{array}{c}
0 \\ 
0 \\ 
0 \\ 
1%
\end{array}%
\right) .
\end{eqnarray}%
The output state is the product state $\left| 1\right\rangle _{1}\left|
1\right\rangle _{2}$. \ 

The optical device that performs the analogous transformation is shown in
Fig. 2b. \ The device, again, uses half-wave plates to implement the
Hadamard transformations, and the conditional-phase switch which is set to
contribute an amplitude of +$\varepsilon $ for a horizontally-polarized
photon pair. \ The input state to this device, $\left| \psi
_{1}\right\rangle ,$ is again described by the general down-conversion state,%
\begin{equation}
\left| \psi _{1}\right\rangle =\left| 0\right\rangle +\varepsilon \left( 
\begin{array}{c}
\alpha \\ 
\beta \\ 
\gamma \\ 
\delta%
\end{array}%
\right) .
\end{equation}%
This state passes through the polarization rotator in mode 1 and will evolve
to the state $\left| \psi _{2}\right\rangle $,%
\begin{eqnarray}
\left| \psi _{2}\right\rangle &=&\left( H_{1}\otimes I_{2}\right) \left|
\psi _{1}\right\rangle \\
&=&\left| 0\right\rangle +\frac{\varepsilon }{\sqrt{2}}\left[ 
\begin{array}{cccc}
1 & 0 & 1 & 0 \\ 
0 & 1 & 0 & 1 \\ 
1 & 0 & -1 & 0 \\ 
0 & 1 & 0 & -1%
\end{array}%
\right] \left( 
\begin{array}{c}
\alpha \\ 
\beta \\ 
\gamma \\ 
\delta%
\end{array}%
\right) \\
&=&\left| 0\right\rangle +\frac{\varepsilon }{\sqrt{2}}\left( 
\begin{array}{c}
\alpha +\gamma \\ 
\beta +\delta \\ 
\alpha -\gamma \\ 
\beta -\delta%
\end{array}%
\right) .
\end{eqnarray}%
This state is subsequently passed through the conditional-phase switch where
the pump laser is set to the appropriate amplitude and phase to add an
amplitude of $+\varepsilon $ for a vertically-polarized photon pair. \ The
state evolves to $\left| \psi _{3}\right\rangle $ where%
\begin{eqnarray}
\left| \psi _{3}\right\rangle &=&\left| 0\right\rangle +\frac{\varepsilon }{%
\sqrt{2}}\left( 
\begin{array}{c}
\alpha +\gamma \\ 
\beta +\delta \\ 
\alpha -\gamma \\ 
\beta -\delta%
\end{array}%
\right) +\varepsilon \left( 
\begin{array}{c}
1 \\ 
0 \\ 
0 \\ 
0%
\end{array}%
\right) \\
&=&\left| 0\right\rangle +\frac{\varepsilon }{\sqrt{2}}\left( 
\begin{array}{c}
\alpha +\gamma +\sqrt{2} \\ 
\beta +\delta \\ 
\alpha -\gamma \\ 
\beta -\delta%
\end{array}%
\right) .
\end{eqnarray}%
Finally, this state passes through a pair of half-wave plates. \ The final
state, $\left| \psi ^{\prime }\right\rangle $, is%
\begin{eqnarray}
\left| \psi ^{\prime }\right\rangle &=&\left| 0\right\rangle +\frac{1}{2}%
\left[ 
\begin{array}{cccc}
1 & 1 & 1 & 1 \\ 
1 & -1 & 1 & -1 \\ 
1 & 1 & -1 & -1 \\ 
1 & -1 & -1 & 1%
\end{array}%
\right] \frac{\varepsilon }{\sqrt{2}}\left( 
\begin{array}{c}
\alpha +\gamma +\sqrt{2} \\ 
\beta +\delta \\ 
\alpha -\gamma \\ 
\beta -\delta%
\end{array}%
\right) \\
&=&\left| 0\right\rangle +\sqrt{2}\varepsilon \left( 
\begin{array}{c}
\alpha +\beta +\frac{1}{\sqrt{2}} \\ 
\alpha -\beta +\frac{1}{\sqrt{2}} \\ 
\gamma +\delta +\frac{1}{\sqrt{2}} \\ 
\gamma -\delta +\frac{1}{\sqrt{2}}%
\end{array}%
\right) .
\end{eqnarray}%
If, for example our input state has $\alpha =\delta =-1/\sqrt{2}$ and $\beta
=\gamma =0$ (i.e. the input is $\left| 0\right\rangle -\varepsilon \left|
\phi ^{+}\right\rangle $ -- one of the outputs of the previous device), then
the output state would be,%
\begin{eqnarray}
\left| \psi ^{\prime }\right\rangle &=&\left| 0\right\rangle +\sqrt{2}%
\varepsilon \left( 
\begin{array}{c}
0 \\ 
0 \\ 
0 \\ 
\sqrt{2}%
\end{array}%
\right) \\
&=&\left| 0\right\rangle +\varepsilon \left( 
\begin{array}{c}
0 \\ 
0 \\ 
0 \\ 
1%
\end{array}%
\right) .
\end{eqnarray}%
That is, the output contains only an amplitude for a photon pair in the
product state $\left| V\right\rangle _{1}\left| V\right\rangle _{2}$. \ The
results for all of the input states are simply stated:%
\begin{eqnarray}
\left| 0\right\rangle -\varepsilon \left| \psi ^{-}\right\rangle
&\longrightarrow &\left| 0\right\rangle +\varepsilon \left| H\right\rangle
_{1}\left| H\right\rangle _{2}  \notag \\
\left| 0\right\rangle -\varepsilon \left| \psi ^{+}\right\rangle
&\longrightarrow &\left| 0\right\rangle +\varepsilon \left| H\right\rangle
_{1}\left| V\right\rangle _{2}  \notag \\
\left| 0\right\rangle -\varepsilon \left| \phi ^{-}\right\rangle
&\longrightarrow &\left| 0\right\rangle +\varepsilon \left| V\right\rangle
_{1}\left| H\right\rangle _{2}  \notag \\
\left| 0\right\rangle -\varepsilon \left| \phi ^{+}\right\rangle
&\longrightarrow &\left| 0\right\rangle +\varepsilon \left| V\right\rangle
_{1}\left| V\right\rangle _{2},
\end{eqnarray}%
and are the inverse of the transformation the previous device
performed.\qquad

In order to complete the measurement of the Bell state, the output of this
device is passed through an optical device like the one in Fig. 3. \ The
detection of a photon pair constitutes a successful measurement and will
occur with probability $\left| \varepsilon \right| ^{2}$ -- the probability
of having a Bell state in our input state. \ This probability ignores issues
of detector and path efficiency.

\section{Discussion}

We have proposed a way of implementing a transformation capable of
converting the polarization state of a pair of photons from the rectilinear
basis to the Bell state basis and vice versa provided the photon pairs are
in a known coherent superposition with the vacuum. \ This transformation
relies on a recently reported effective nonlinearity at the single-photon
level \cite{phase}. \ Requiring the photon pair to be in a superposition
with the vacuum seems unusual, but this type of superposition exists in all
down-conversion\ sources of entangled photons. \ It is only upon performing
a photon-counting coincidence measurement that the maximally-entangled
behaviour is projected out. \ While these down-conversion\ sources of Bell
states exist and are practical in the lab, the creation mechanism does not
suggest how one might try to measure those Bell states. \ In the device
discussed here, the Bell state creator and Bell state analyzer look very
similar. \ The creator can essentially be run in reverse to make the
analyzer. \ 

This device cannot be used for performing unconditional quantum
teleportation. \ The device is only capable of distinguishing the four Bell
states; it is not capable of performing a general projective measurement in
the Bell basis. \ This is due to the conditional-phase shifter's dependence
on the magnitude and phase of the amplitude for the Bell state component in
the input state; the gate does not operate properly on arbitrary
superpositions of Bell states.\ Nevertheless, the device discussed herein
constitutes a novel way of manipulating the degree of entanglement between a
pair of photons, and may find a use in other quantum optics applications,
such as dense coding \cite{densetheory,denseexp}. \ The ability to entangle
and disentangle photon pairs is a crucial step toward building scalable
all-optical quantum computers.

We would like to thank Andrew White and Ray Laflamme for valuable
discussions. \ We are grateful for the financial support of Photonics
Research Ontario, NSERC, and the US Air Force Office of Scientific Research
(F49620-01-1-0468).

\textbf{Fig. 1.\ a) A quantum circuit and b) its optical analogue for the
creation of Bell states from product states. \ a) The quantum circuit acts
on a pair of input modes 1 and 2. \ The circuit uses one-qubit Hadamard
gates, and a two-qubit controlled-}$\pi $\textbf{\ gate. \ This circuit
performs a unitary transformation on the inputs and takes each of the four
possible qubit product states to a different Bell state. \ b) The optical
analogue of the quantum circuit. \ In the diagram, }$\lambda /2$ \textbf{are
half-wave plates oreinted at 22.5 degrees and }$\chi ^{(2)}$ \textbf{is a
nonlinear material. \ \ The device is capable of converting the state of a
photon pair in a product state of polarization to one of the Bell states,
provided that the input is in the correct superposition with the vacuum. \ }

\textbf{Fig. 2. a) A quantum circuit and b) its optical analogue for the
conversion of Bell states to product states. \ a) This quantum circuit takes
a pair of qubits in input modes 1 and 2 and performs a unitary
transformation that will convert a Bell state to a product state. \ b) The
optical analogue of the quantum circuit takes a photon pair in a Bell state
to a rectilinear product state, provided the photon pair is in the correct
superposition with the vacuum. }

\textbf{\ \ }

\textbf{Fig. 3. An optical device for distinguishing rectlinear basis
states. \ This simple device can distinguish between the product states for
the polarization of a pair of photons }$\left| H\right\rangle _{1}\left|
H\right\rangle _{2}$\textbf{, }$\left| H\right\rangle _{1}\left|
V\right\rangle _{2}$\textbf{, }$\left| V\right\rangle _{1}\left|
H\right\rangle _{2}$\textbf{, and }$\left| V\right\rangle _{1}\left|
V\right\rangle _{2}$\textbf{, where the subscripts 1 and 2 are mode labels.
\ The device consists of a pair of polarizing beam-splitters\ (PBS) and 4
photon counting detectors monitoring their outputs. \ For example, the
detection of a photon at detector 1 and detector 4 corresponds to the state }%
$\left| H\right\rangle _{1}\left| V\right\rangle _{2}$\textbf{.}

\textbf{Fig. 4. Schematic for the conditional-phase switch. \ A strong,
classical, laser in mode p, of frequency 2}$\omega $\textbf{,} \textbf{pumps
a }$\chi ^{(2)}$ \textbf{nonlinear material such that it can create
down-conversion pairs in modes 1 and 2. \ A pair of input beams, of
frequency }$\omega $\textbf{, pass through the nonlinear material into modes
1 and 2. \ Interference between the multiple paths leading to photon pairs
at the output can be used to introduce a large phase shift on the amplitude
for a photon pair.}

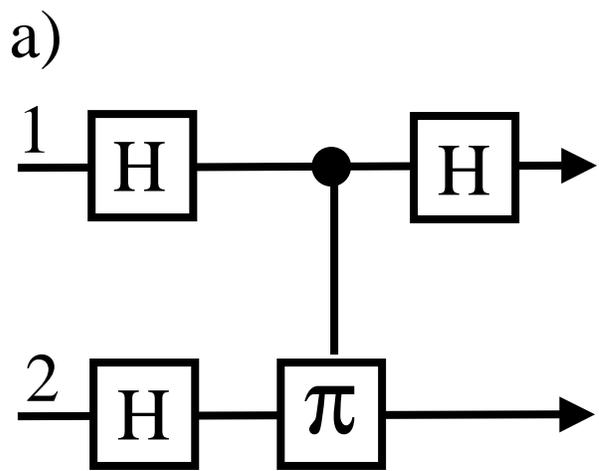
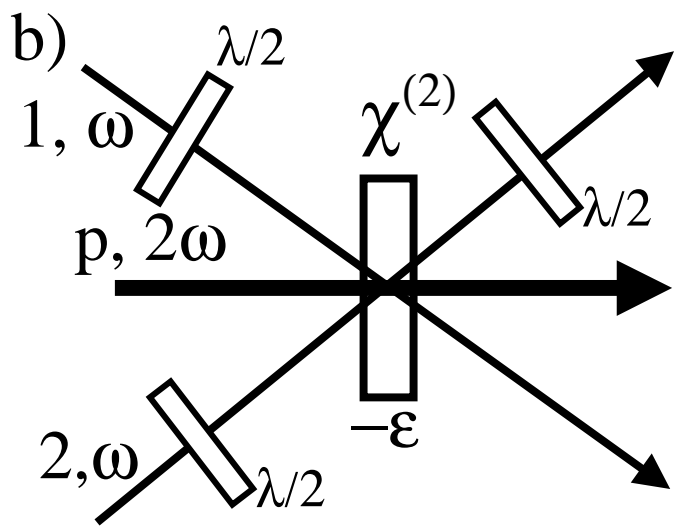
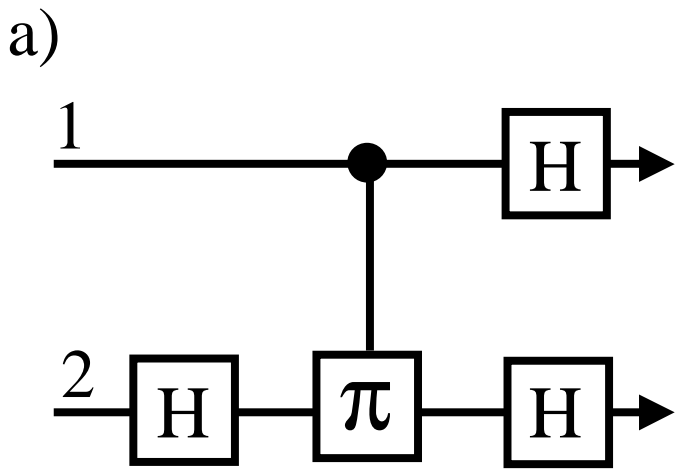
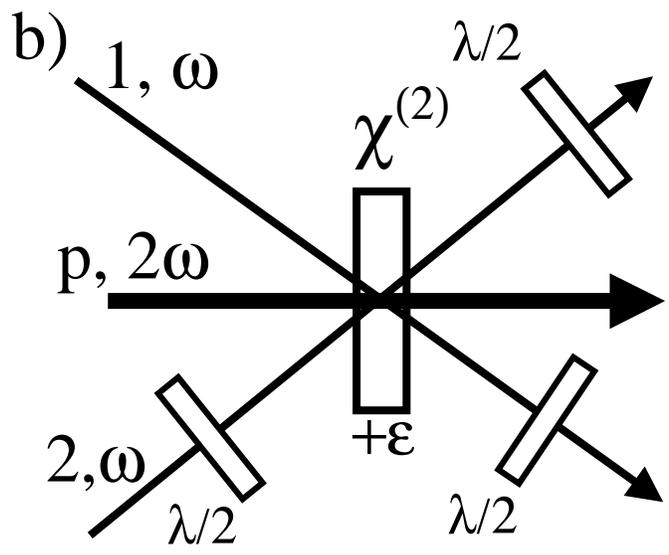
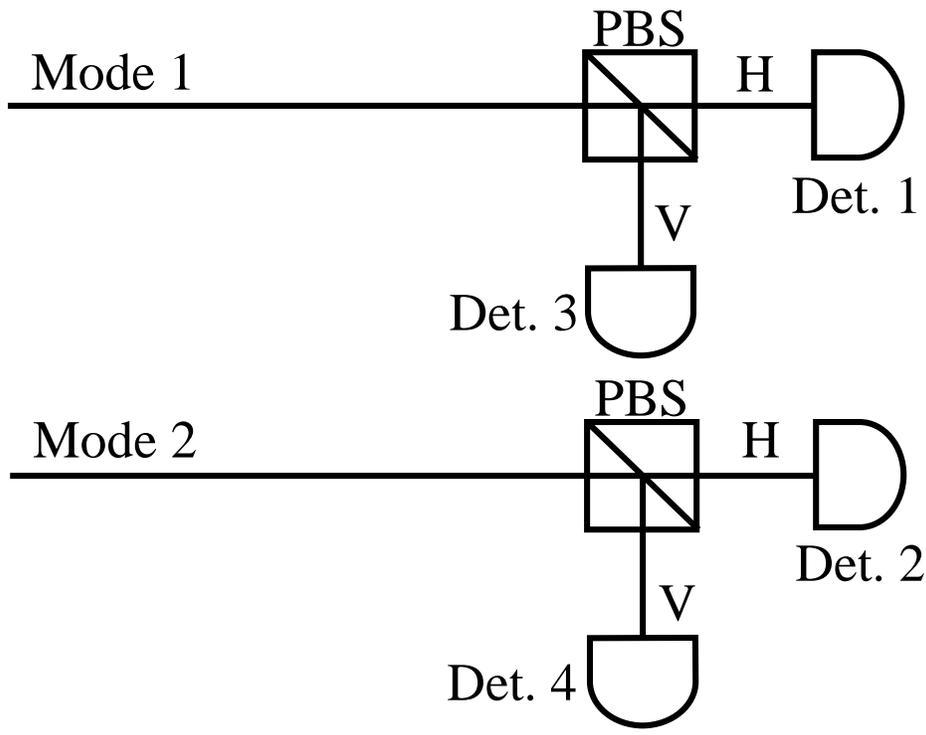
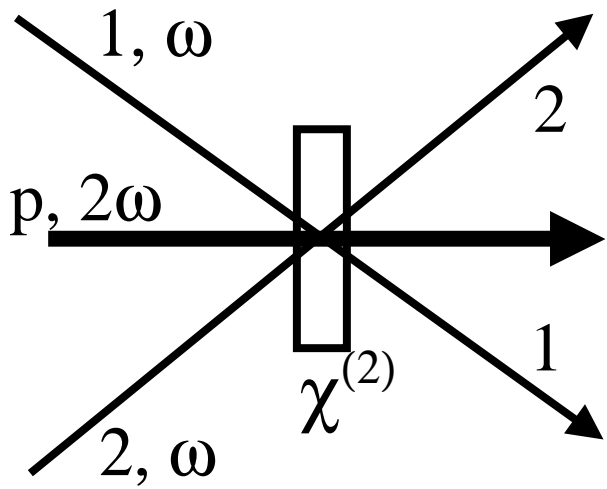
\end{document}